\documentclass[aps,pra,twocolumn]{revtex4}

\usepackage{amsfonts}
\usepackage{amsmath}
\usepackage{graphicx}
\usepackage{dsfont}
\usepackage{diagbox}
\usepackage{amssymb}
\usepackage{bbm}
\usepackage{float}
\usepackage{subfigure}
\usepackage{url}
\usepackage{hyperref}
\usepackage{multirow}
\usepackage{bm}
\usepackage{booktabs}
\usepackage{array}
\usepackage{color}
\usepackage{longtable}
\usepackage{booktabs}
\usepackage{algorithm}
\usepackage{algorithmic}

\newcommand{\PreserveBackslash}[1]{\let\temp=\\#1\let\\=\temp}
\newcolumntype{C}[1]{>{\PreserveBackslash\centering}p{#1}}
\newcolumntype{R}[1]{>{\PreserveBackslash\raggedleft}p{#1}}
\newcolumntype{L}[1]{>{\PreserveBackslash\raggedright}p{#1}}
\newcommand{\marked}[1]{{\color{black}{#1}}}

\begin{document}

\title{Neural canonical transformations for vibrational spectra of molecules}

\author{Qi Zhang$^{1}$, Rui-Si Wang$^{1,2}$, Lei Wang$^{1,3}$}
\affiliation{$^{1}$ Beijing National Laboratory for Condensed Matter Physics and Institute of Physics, 
Chinese Academy of Sciences, Beijing 100190, China \\
$^2$University of Chinese Academy of Sciences, Beijing 100049, China \\
$^3$Songshan Lake Materials Laboratory, Dongguan, Guangdong 523808, China}
\date{\today}
	
\begin{abstract}

	The behavior of polyatomic molecules around their equilibrium positions can be regarded as quantum-coupled anharmonic oscillators.
	Solving the corresponding Schrödinger equations can interpret or predict experimental spectra of molecules.
	In this study, we develop a novel approach to solve excited states of anharmonic vibrational systems.
	The normal coordinates of molecules are transformed into new coordinates through a normalizing flow parameterized by a neural network,
	facilitating the construction of a set of orthogonal many-body variational wavefunctions.
	This methodology has been validated on an exactly solvable $64$-dimensional coupled harmonic oscillator, 
	yielding numerical results with a relative error on the order of $10^{-6}$. 
	Furthermore, the neural canonical transformations are also applied to calculate the energy levels of two specific molecules, acetonitrile ($\text{C}\text{H}_3\text{CN}$) and 
	ethylene oxide ($\text{C}_2\text{H}_4\text{O}$) involving $12$ and $15$ vibrational modes respectively, 
	which are in agreement with experimental findings.
	One of the key advantages of this approach is its flexibility concerning the potential energy surface,
	requiring no specific form for its application.
	Moreover, this method can be easily implemented on large-scale distributed computing platforms, 
	making it particularly suitable for investigating more complex vibrational structures.

\end{abstract}
\maketitle
\section{Introduction}

Determining the vibrational frequencies of molecules plays a crucial role in investigating their structure and physical properties.
The most direct experimental methods involve utilizing infrared absorption spectroscopy and Raman spectroscopy.
However, obtaining the entire spectrum for molecules with a large number of vibration modes through experimentation presents a significant challenge.
In such instances, the reliance on theoretical results becomes necessary for prediction and interpretation.
To effectively capture and describe the properties of these molecules, 
we can consider the interactions within them as vibrational systems.
It has been proved that using anharmonic quantum oscillators is a convenient and efficient approximation for studying molecules\cite{bowman1978self,bowman1986self,gerber1979semiclassical,thompson1982optimization,christoffel1982investigations,rauhut2007configuration,scribano2008iterative,gohaud2005new,begue2006single,begue2007comparison,cassam2003alternative,cassam2003ab,manzhos2009using,yagi2012optimized,ishii2022development,leclerc2014calculating,thomas2015using,rakhuba2016calculating,knyazev2001toward,odunlami2017vci,saleh2023computing,bowman2008variational,zoccante2012approximate,godtliebsen2013tensor,madsen2018tensor,madsen2021calculating,erba2019anharmonic}.
However, the complex coupling interactions in the Hamiltonian result in the Schrödinger equations that are not exactly solvable.
This inherent complexity poses a particularly challenging obstacle when striving to calculate numerous energy levels accurately. 
Hence, there has been significant interest in developing new numerical methods to calculate the excited state energies of vibrational systems.

Over the past decades, various numerical methods have been developed to solve the excited states of vibrational Hamiltonian for anharmonically coupled oscillators.
Some approaches are based on the quantum self-consistent field theory, 
such as the vibrational self-consistent-field (VSCF) method~\cite{bowman1978self,bowman1986self,gerber1979semiclassical,thompson1982optimization},
which represents the variational wavefunction as a Hartree product,
and it neglects the interactions between different modes.
To address this issue, the vibrational configuration interaction (VCI) method~\cite{christoffel1982investigations,rauhut2007configuration,scribano2008iterative} was developed,
which involves the quantum correlations with different modes and provides more accurate results.
Some improvements have been adopted to generate other variational methods, 
such as parallel vibrational multiple window configuration interaction (P-VMWCI)~\cite{gohaud2005new,begue2006single,begue2007comparison}, the vibrational mean field configuration interaction (VMFCI)~\cite{cassam2003alternative,cassam2003ab,begue2007comparison},
the adaptive VCI~\cite{odunlami2017vci},
optimized-coordinate VSCF, and optimized-coordinate VCI~\cite{yagi2012optimized} methods.
With the development of the tensor network,
including the reduced-rank block power method (RRBPM)~\cite{leclerc2014calculating,thomas2015using} and the locally optimal block preconditioned conjugate gradient (LOBPCG) method~\cite{rakhuba2016calculating,knyazev2001toward}. 
However, to accurately measure high entanglement states within a finite memory, it is necessary to use special designs for the tensor network structures. 
Meanwhile, a series of vibrational coupled cluster (VCC)~\cite{zoccante2012approximate,godtliebsen2013tensor,madsen2018tensor,madsen2021calculating} methods were developed, 
where the reference wavefunctions are obtained by VSCF, 
and then parameterization it nonlinearly.
These are capable of providing higher precisions.
It is necessary to represent the potential energy surface (PES) in the form of a sum of products for all tensor-based methods.

The neural canonical transformation has proven to be a successful tool in solving many-body questions, such as interacting electron gas and dense hydrogen~\cite{witriol1972canonical,xie2022abinitio,xie2023mstar,xie2023deep}.
This newly developed technique utilizes neural networks to parameterize the normalizing flow~\cite{dinh2014nice,germain2015made,rezende2016variational,dinh2016density,papamakarios2019neural,papamakarios2021normalizing,saleh2023computing,wang2018generatemodels}.
Normalizing flow is a concept used in machine learning and generative modeling to transform a simple probability distribution into a more complex one while ensuring that the transformed distribution is normalized.
The transform is reversible, meaning that the transformation can be seen as a bijection between two probability distributions. 
This capability makes it possible to parameterize a set of complex wavefunctions of anharmonic oscillators from harmonic ones, which is useful for studying vibrational systems.

The neural canonical transformation diverges from conventional neural network strategies,
which only focus on calculating a limited number of low-lying excited states~\cite{manzhos2009using}.
In a recent study~\cite{saleh2023computing}, a normalizing flow was employed for computing excited states.
They use a non-orthogonal basis, which is a significant computational effort for calculating the integrals of energies in multiple dimensions and requires additional diagonalization operations,
which limits its applicability to more extensive applications.
In Ref.~\cite{ishii2022development}, the authors use a backflow transformation to solve the excited states. 
However, they do not ensure orthogonality among various states and measure the energies directly, which violates the variational principle.
In another recent work~\cite{pfau2023natural}, 
researchers study the molecules by utilizing the Born-Oppenheimer approximation, 
and they use neural networks to create a set of non-orthogonal variational wavefunctions for electrons,
requiring an extra diagonalization and more computational resources, scaling at $\mathcal{O}(M^4)$ for $M$ states.
As a result, they were only able to calculate a few low-lying states with several electrons.
Furthermore, in Ref.~\cite{cranmer2019inferring}, the authors characterize an anharmonic oscillator using a density matrix to study its lowest excited states via normalizing flow but focus only on a single-particle system.

In this study, 
we attempt to provide a novel method for solving anharmonic vibrational systems,
i.e., the neural canonical transformation approach.
This approach involves the formulation of a set of orthogonal variational wavefunctions by constructing a bijective mapping with non-interacting wavefunctions.
The main advantages include orthogonality of all excited states under normalizing flow, flexibility in potential forms, precise calculation of higher excited state energies, and support for parallel computations.
This enhances computational efficiency and facilitates extension to large polyatomic molecules with complex interactions.
In practice, 
the real non-volume preserving (RNVP)~\cite{dinh2016density} neural network is employed to parameterize the normalizing flow~\cite{dinh2014nice,germain2015made,rezende2016variational,dinh2016density,papamakarios2019neural,papamakarios2021normalizing,saleh2023computing,wang2018generatemodels}.
Markov chain Monte Carlo (MCMC)~\cite{Becca2017} is utilized for sampling the coordinates
and automatic differentiation is used for computing the gradient of the loss function.
To validate this approach,
we compute the energy levels of a 64-dimensional coupled harmonic oscillator~\cite{rakhuba2016calculating,leclerc2014calculating,thomas2015using} and two actual molecules, 
acetonitrile ($\text{C}\text{H}_3\text{CN}$) with $12$ variational modes~\cite{Herzberg1960MolecularSA,nakagawa1962rotation,Koivusaari1992high,tolonen1993the,andrews2000vibrational,paso1994the,duncan1978methyl,nakagawa1985infrared,begue2005calculations,thomas2015using,leclerc2014calculating,avila2011using,odunlami2017vci},
and ethylene oxide ($\text{C}_2\text{H}_4\text{O}$) with $15$ modes~\cite{komornicki1983ab,lowe1986scaled,rebsdat2000ethylene,begue2006single,cant1975vibrational,lord1956vibrational,schriver2004vibrational,nakanaga1980coriolis,yoshimizu1975microwave,carbonniere2010vci, zoccante2012approximate, thomas2015using} as a benchmark test.
Moreover, our program is written in Python with JAX~\cite{jax2018github}. 
All the source codes, training data of neural networks, and raw data of energies are available in the open-access repository~\cite{mycode}.

\section{Method}\label{sec:method}

For an actual molecule, we can apply the Born-Oppenheimer approximation to separate the system into vibrational and electronic components. 
The electronic part can be described by potential energy surfaces (PES), which are generated using methods like density functional theory (DFT) and coupled cluster single and double excitations (CCSD).
In the vibrational Hamiltonian, we use the Watson kinetic energy operator (KEO)~\cite{Nielsen1951, James1968}
and only focus on pure vibrational calculation.
\marked{Also, the vibrational angular momentum terms are neglected in this work, 
which will introduce slight modifications to the vibration energies~\cite{bowman2008variational}.
These neglected terms may cause some differences between the numerical and the experimental results.}
Then,
the system can be characterized by a Hamiltonian in $D$ vibrational degrees of freedom in the normal coordinates $\bm{x}$:
\begin{equation}
  H = 
  - \frac{1}{2} \sum_{i=1}^D  \frac{\partial^2}{\partial x_i^2} 
  + V(\bm{x})
  =  \frac{1}{2} \sum_{i=1}^D \left(
	- \frac{\partial^2}{\partial x_i^2} 
	+ \omega_i^2 x_i^2
  \right)
  + V_\text{anh}(\bm{x}),
\end{equation}
where $\bm{x}=(x_1,x_2,...,x_D)$ spans the $D$-dimensional real space $\mathcal{R}^D$, 
and each $x_i$ denoting the coordinate associated with a distinct vibrational mode.
The term $V(\bm{x})$ represents the PES,
which incorporates both the harmonic components and the more intricate interactions.

\begin{figure}[h]
	\centering
	\includegraphics[width=0.28\textwidth]{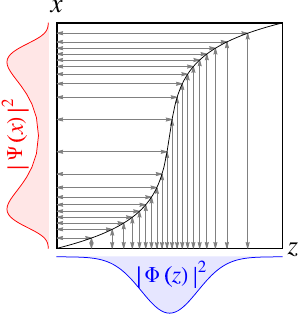}
	\caption{A sketch of a normalizing flow ($f: x\rightarrow z$) from a density distribution $|\Psi(x)|^2$ to another density distribution $|\Phi(z)|^2$. 
	It is implemented through a reversible coordinate transformation.
	In functional form, it can be expressed as $z = f_{\theta}(x)$ and $x = f^{-1}_{\theta}(z)$, where $\theta$ is the parameters of the flow.}
	\label{Fig:Flow}
\end{figure}

The wavefunctions associated with the vibrational Hamiltonian are denoted by $| \Psi_n\rangle$, 
where the index $n$ distinct energy levels, and notably, $n=0$ identifies the ground state. 
These eigenstates constitute a set of orthogonal many-body states, 
with each state corresponding to an eigenvalue $E_n = \langle \Psi_n |H| \Psi_n \rangle$. 
To compute these wavefunctions using the variational method, choosing a good parameterization is crucial.
A natural approach involves applying a unitary transformation $U$ to the non-interacting reference states $| \Phi_n\rangle$,
which are the wavefunctions for $D$-dimensional harmonic oscillators. 
We can use the normalizing flow~\cite{dinh2014nice,germain2015made,rezende2016variational,dinh2016density,papamakarios2019neural,papamakarios2021normalizing,saleh2023computing,wang2018generatemodels} to make a bijection between these two set of wavefunctions.
It can be realized via a parameterized and learnable bijection that maps the normal coordinates $\bm{x}$ to a set of new coordinates $\bm{z}$ in $D$-dimensional space,
which can be understood as a smooth, reversible function $\bm{z} = f_{\bm{\theta}}(\bm{x})$ and $\bm{x} = f^{-1}_{\bm{\theta}}(\bm{z})$.
Typically, the function $f$ is composed of neural networks,
and the associated parameters in the networks are denoted as $\bm{\theta}$.
The normalizing flow serves as an elaboration of the backflow transformation through the application of invertible neural networks, offering a modern approach to handling transformations.
We also represent a normalizing flow from a complex distribution to a simple Gaussian distribution in FIG. \ref{Fig:Flow}, 
which shows that the flow establishes a bijection between two distinct distributions.
The many-body wavefunctions are then articulated as follows~\cite{xie2022abinitio,xie2023mstar,xie2023deep,saleh2023computing}:
\begin{equation}\label{Eq:wavefunction}
	\Psi_{n}(\bm{x}) = \Phi_n(f_{\bm{\theta}}(\bm{x})) 
	\left|\text{det}\left( \frac{\partial f_{\bm{\theta}}(\bm{x})}{\partial \bm{x}} \right)\right|^{\frac{1}{2}},
\end{equation}
or it can be abbreviated as $\Psi_{n}(\bm{x})=\Phi_{n}(\bm{z})\left|\text{det}\left( {\partial \bm{z}}/{\partial \bm{x}} \right)\right|^{1/2}$,
where $\Psi_{n}(\bm{x}) = \langle \bm{x}| \Psi_n \rangle$, and $\Phi_{n}(\bm{z})= \langle \bm{z}| \Phi_n \rangle$.
The variational wavefunction is dependent on parameters $\bm{\theta}$ and the basis states $\Phi_{n}(\bm{z})$.
Usually, the basis state can be chosen as a Hartree product of harmonic oscillators,
which are detailed in Appendix \ref{sec:basisstate}.

The unitary transformation, generated by the square root of the Jacobian determinant
$U_{\bm{\theta}}(\bm{x}) = \left|\text{det}\left( {\partial f_{\bm{\theta}(\bm{x})}}/{\partial \bm{x}} \right)\right|^{1/2}$,
plays a critical role in introducing correlations among various vibrational modes, 
thereby effectively approximating complicated many-body wavefunctions.
This transformation is also known as a point transformation, as delineated in Refs.~\cite{dewitt1952point,eger1963point,eger1964point,witriol1972canonical}. 
The use of a set of orthogonal basis states $\Phi_{n}(\bm{z})$ alongside the square root of the Jacobian determinant is pivotal in this method.
This mathematical structure ensures that the transformed wavefunctions $\Psi_{n}(\bm{x})$ maintain their orthogonality~\cite{xie2022abinitio,xie2023mstar,xie2023deep,saleh2023computing}:
\begin{equation}
	\int  \Psi^*_{n'}(\bm{x})\Psi_{n}(\bm{x})  d\bm{x} = \delta_{nn'}.
\end{equation}

It is worth emphasizing that this variational wavefunction markedly diverges from those traditionally used in the VSCF method~\cite{bowman1978self,bowman1986self,gerber1979semiclassical,thompson1982optimization}. 
Our formulation incorporates an additional Jacobian determinant, 
substantially amplifying the expressive capabilities of the wavefunction.
This modification not only augments the theoretical framework but also enhances the potential for capturing intricate interactions in vibrational systems.

Next, 
we focus on optimizing the variational wavefunctions for all target states,
including both the ground and excited states. 
The primary goal is to determine the optimal coefficients $\bm{\theta}$ for the normalizing flow $f$.
Within this framework, the loss function can be succinctly defined as the summation of energies across all states:
\begin{equation}\label{Eq:lossfunction}
	\mathcal{L} = \sum_{n=0}^{M-1} E_{n},
\end{equation}
with
\begin{equation}\label{Eq:lossfunction}
	E_{n}
	= \int \Psi_n^*(\bm{x}) H \Psi_n(\bm{x}) d\bm{x}
	\approx \underset{\bm{x}\sim |\Psi_{n}(\bm{x})|^2}{\large{\mathbb{E}}} 
	\left[ E_{n}^{\text{loc}}(\bm{x}) \right],
\end{equation}
where $M$ denotes the number of states targeted for computation.
Because high-dimensional integrals require significant computational resources,
we use the Markov chain Monte Carlo (MCMC)~\cite{Becca2017} sampling method to approximate the energy of each state $E_n$.
For a given state $n$, 
the coordinates are sampled according to the probability distribution $|\Psi_{n}(\bm{x})|^2$, and the number of samples is called the batch size $B$.
Using $j~(j=1, 2, ..., B)$ to label these samples, resulting in a series of coordinates $\bm{x}^j$.
The energy of sample with coordinates $\bm{x}^j=(x_1^j,x_2^j,...,x_D^j)$ is called the local energy $E_{n}^{\text{loc}}(\bm{x}^j)$
and we can calculate the average of these samples to approximate the expectation energy of state $E_n \approx \sum_{j=1}^B E_{n}^{\text{loc}}(\bm{x}^j) /B$,
a process typically denoted by the expectation symbol 
${\large{\mathbb{E}}}_{\bm{x}\sim |\Psi_{n}(\bm{x})|^2} 
\left[ E_{n}^{\text{loc}}(\bm{x}) \right]$.
The above local energies are associated with the wavefunctions~\cite{xie2022abinitio,xie2023mstar}:
\begin{equation}\label{Eq:localenergy}
	\begin{aligned}
	&E_{n}^{\text{loc}}(\bm{x})
	 = \frac{H \Psi(\bm{x})}{\Psi(\bm{x})} \\
	&~~~~= - \frac{1}{2} \sum_{i=1}^D \left[
	\frac{\partial^2}{\partial x_i^2} \ln |\Psi_{n}(\bm{x})| 
	+ \left( \frac{\partial}{\partial x_i}  \ln |\Psi_{n}(\bm{x})|  \right)^2
	\right] + V(\bm{x}),
	\end{aligned}
\end{equation}
where we take logarithms of the wavefunctions to avoid the numerical problem in computing, the details are presented in Appendix \ref{sec:localenergy}.
The gradient and laplacian of the logarithm wavefunction, crucial for measuring the kinetic energy, can be computed efficiently through automatic differentiation.
This technique, supported by popular computational packages such as JAX~\cite{jax2018github} and PyTorch~\cite{paszke2019pytorch}, offers significant advantages in terms of speed and accuracy over traditional numerical differentiation methods.
Furthermore, this method exhibits linear computational complexity $\mathcal{O}(M)$ dependent on the number of state samplings, as demonstrated by Eq. (\ref{Eq:lossfunction}).

Crucially, our approach eliminates the reliance on the specific form of the PES,
which can be represented in various forms, including series expansions, neural networks, or even DFT programs. 
\marked{This flexibility sets the approach apart from other methods such as VSCF and VCI, as well as some tensor-based methods like RRBPM and LOBPCG, which usually require the PES to be defined as a sum of products to avoid expensive numerical integrals. 
Recent advances in machine learning have led to the creation of PES through neural networks~\cite{lu2022fast,alvertis2023phonon}, providing more options for PES construction.
Hence, the neural canonical transformation provides enhanced flexibility in selecting PES, thereby ensuring broader applicability in modeling complex vibrational systems.}

In the training procedure,
the optimization of parameters is essential through the gradient of the loss function,
which can be derived from Eq. (\ref{Eq:lossfunction}) and Eq. (\ref{Eq:localenergy}).
It is also referred in the Refs.~\cite{xie2022abinitio,xie2023mstar}:
\begin{equation}
	\nabla_{\bm{\theta}} \mathcal{L} = 2 \sum_{n=0}^M
	\underset{\bm{x}\sim |\Psi_{n}(\bm{x})|^2}{\large{\mathbb{E}}}
	\left[
		E_{n}^{\text{loc}}(\bm{x})~
		\nabla_{\bm{\theta}} \ln |\Psi_{n}(\bm{x})|
	\right].
\end{equation}
Then we can optimize the parameters $\bm{\theta}$ through the well-known adaptive moment estimation (ADAM)~\cite{kingma2014adam} optimizer, which is commonly used in machine learning.
It can be understood as $\bm{\theta}_{\text{new}} = \bm{\theta} - \alpha ~\nabla_{\bm{\theta}} \mathcal{L}$, where $\alpha$ is the learning rate, set to $\alpha=0.001$ in this work.
In summary, the entire process of the neural canonical transformation approach for determining the energy levels of vibrational systems is outlined in TABLE~\ref{alg}.

\begin{table}
	\begin{tabular}{L{8.6cm}}
		\hline\hline \\[-6.0ex]
	\caption{Algorithm: Energy levels in vibrational systems}\label{alg}\\[-3.0ex]
		\hline \\[-4.0ex]
	\begin{algorithmic}[1]
		\item[\textbf{Input:}] Potential energy surface $V(\bm{x})$, batch size $B$, learning rate $\alpha$.
		\item[\textbf{Output:}] Energy levels $E_n$, neural network parameters $\bm{\theta}$.
	\STATE Initialize the variational wavefunctions $\Psi_{n}(\bm{x}) = \Phi_n(f_{\bm{\theta}}(\bm{x})) 
	\sqrt{
		\left|\text{det}\left( {\partial f_{\bm{\theta}}(\bm{x})}/{\partial \bm{x}} \right)\right|}$
	\WHILE{$\mathcal{L}$ is not converged}
	\FOR{$n=0,1,2,...,M-1$}
		\STATE ~~~~MCMC Sampling $\bm{x}\sim |\Psi_n(\bm{x})|^2$ with batch size $B$, thus obtaining $\bm{x}^j~~(j=1,2,..., B)$.
		\STATE ~~~~Measure the local energy $E_{n}^{\text{loc}}(\bm{x}^j)$ of each $\bm{x}^j$.
		\STATE ~~~~Compute energy levels $E_n \approx \sum_{j=1}^B E_{n}^{\text{loc}}(\bm{x}^j)/B$.
	\ENDFOR
	\STATE Compute the loss function $\mathcal{L}=\sum_{n=0}^{M-1} E_{n}$ and the gradient of loss $\nabla_{\bm{\theta}} \mathcal{L}$. 
	\STATE Optimize parameters $\bm{\theta}$ via ADAM optimizer, i.e. $\bm{\theta}_{\text{new}} = \bm{\theta} - \alpha~\nabla_{\bm{\theta}} \mathcal{L}$.
	\ENDWHILE
	\end{algorithmic}\\[-2.0ex]
\hline\hline
\end{tabular}
\end{table}

\section{Applications}
\subsection{64-dimensional coupled oscillator}

To evaluate the effectiveness of the method,
we first computed a coupled harmonic oscillator system with $D=64$ dimensions,
serving as a benchmark for testing. 
This system exclusively features quadratic interaction terms, 
making it exactly solvable.
The Hamiltonian of the $D$-dimensional coupled harmonic oscillator is specified as follows~\cite{rakhuba2016calculating,leclerc2014calculating,thomas2015using}:
\begin{equation}\label{Eq:NDCO}
	H = \sum_{i=1}^D 
	\frac{1}{2} \left(- \frac{\partial^2}{\partial x_i^2} 
	+ \omega_i^2 x_i^2 \right)
	+ \sum_{i>j}^{D} \alpha_{ij}\sqrt{\omega_i \omega_j} x_i x_j,
\end{equation}
where we set the frequency for the $i$-th mode as $\omega_i=\sqrt{i/2}$,
and the coupling constant is uniformly chosen as $\alpha_{ij}=0.1$ for all $i>j$,
aligning with the parameters used in prior studies~\cite{rakhuba2016calculating,leclerc2014calculating,thomas2015using}.
\marked{It is noteworthy that the coordinates of this toy model differ from the normal coordinates typically used in molecular, as they include quadratic coupling terms. 
These terms can be eliminated by diagonalizing the Hessian matrix, and a detailed analytical study is provided in Appendix \ref{sec:NDCO}.
However, to validate our method, we perform calculations directly using the coordinates in the Eq. (\ref{Eq:NDCO}).}
The coupling terms result in modifications in both the frequency and excited state energies,
and the primary objective for us is to compute the lowest $M=80$ eigenstates and their corresponding eigenvalues.

In the following computation,
we employed a neural network called RNVP~\cite{dinh2016density}
to parameterize the normalizing flow $f: \bm{x} \rightarrow \bm{z}$,
and a concise introduction to this network is presented in Appendix \ref{sec:RNVP}. 
For this particular benchmark, 
the RNVP network is configured with $16$ coupling layers, 
where each coupling layer consists of a multilayer perceptron (MLP) with $2$ hidden layers and a width of $128$. 
Consequently, the total number of parameters reaches about $0.6$ million.
In the following of this paper, the set of hyperparameters is denoted as $\{16, 128, 2\}$.
A notable advantage of this network is its ability to construct a lower triangular Jacobian determinant,
facilitating the efficient computation of the logarithm of the Jacobian determinant $\log|\partial \bm{z}/\partial \bm{x}|$.

\begin{figure}[h]
	\centering
	\includegraphics[width=0.40\textwidth]{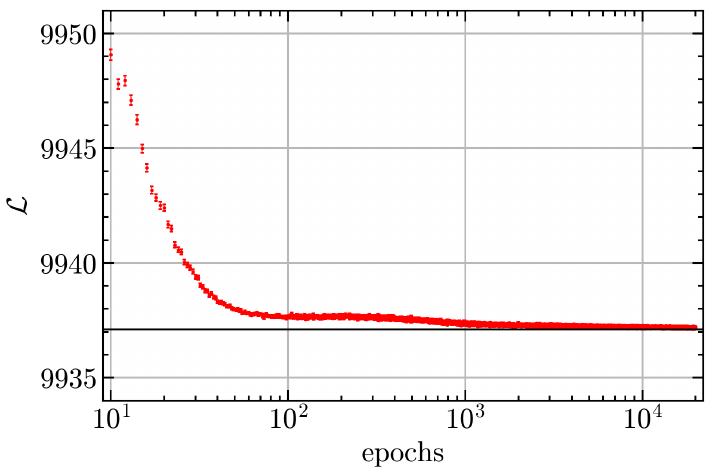}
	\caption{The optimization process of a $64$-dimensional coupled harmonic oscillator.
	The black line represents the values of the analytical solution $\sum_{n=0}^{79} E_{n,\text{exact}}$,
	while the red line depicts the loss function $\mathcal{L}$ with $M=80$ states.}
	\label{Fig:64DCO}
\end{figure}

The training batch size is set to $2048$, 
which indicates that $2048\times 80$ states are sampled through MCMC in each epoch. 
The initial parameters of the network are initialized in proximity to an identity transformation, 
which implies that the initial variational wavefunctions approximate the harmonic oscillator wavefunctions closely. 
FIG. \ref{Fig:64DCO} illustrates the evolution of the loss function with the number of iterations.
After $20000$ epochs, the loss function converges to a value of $\mathcal{L}=9937.186(15)$ energy unit,
which is in close approximation to the analytical solution $\sum_{n=0}^{79} E_{n,\text{exact}}=9937.108$.
Due to the time-consuming nature of a large number of network parameters, 
our tests show that training this network with batch size $B=2048$ and $M=80$ excited states for $20000$ epochs takes about $43$ hours on a single A100 GPU. 
Utilizing multiple GPUs in parallel computation on a cluster can further reduce the computational time,
and we have included code for multi-GPU parallelism in the open-access repository~\cite{mycode}.

\begin{table*}[htp] 
	\caption{A subset of the $M=80$ lowest energy levels of a $64$-dimensional coupled oscillator.
	 In this table, ``$n$" denotes the level index, 
	 ``Assignment" indicates the normal mode assignment, 
	 ``$E_\text{exact}$" represents the exact solution, 
	 ``$E_\text{num}$" corresponds to the numerical results obtained through neural network calculations, 
	 the fifth row provides the relative error,
	 $\Delta E_{n,\text{exact}} = E_{n,\text{exact}} - E_{0,\text{exact}}$ and
	 $\Delta E_{n,\text{num}} = E_{n,\text{num}} - E_{0,\text{num}}$ are excitation energies,
	 and ZPE is an abbreviation for ``zero-point energy", i.e. the ground state energy.
	} 
	\centering
	\begin{tabular}{L{0.8cm}L{2.4cm}R{1.8cm}R{1.8cm}R{1.9cm}R{1.8cm}R{1.8cm}R{2.8cm}}
	  \hline\hline \\[-1.5ex]
	  $n$ & Assignment & $E_{n,\text{exact}}~$ & $E_{n,\text{num}}~$
	  & $\frac{E_{n,\text{num}} - E_{n,\text{exact}}}{E_{n,\text{exact}}}$   
	  & $\Delta E_{n,\text{exact}}$ 
	  & $\Delta E_{n,\text{num}}$  
	  & $\Delta E_{n,\text{num}} - \Delta E_{n,\text{exact}}$
	  \\[1.4ex]
	  \hline \\[-1.5ex]
	  $ 0$ & ZPE & $121.62095$ & $121.62184$ & $7.4\times10^{-6}$ & $-$ & $-$ & $-$~~~~~~ \\
	  $ 1$ & $\nu_{1}$ & $122.29236$ & $122.29311$ & $6.1\times10^{-6}$ & $0.67141$ & $0.67126$ & $-0.00015$~~~~~~ \\
	  $ 2$ & $\nu_{2}$ & $122.58564$ & $122.58644$ & $6.5\times10^{-6}$ & $0.96469$ & $0.96459$ & $-0.00010$~~~~~~ \\
	  $ 3$ & $\nu_{3}$ & $122.81059$ & $122.81153$ & $7.7\times10^{-6}$ & $1.18964$ & $1.18969$ & $0.00005$~~~~~~ \\
	  $ 4$ & $2\nu_{1}$ & $122.96377$ & $122.96457$ & $6.5\times10^{-6}$ & $1.34282$ & $1.34272$ & $-0.00010$~~~~~~ \\
	  $ 5$ & $\nu_{4}$ & $123.00021$ & $123.00141$ & $9.7\times10^{-6}$ & $1.37927$ & $1.37956$ & $0.00030$~~~~~~ \\
	  $ 6$ & $\nu_{5}$ & $123.16727$ & $123.16820$ & $7.5\times10^{-6}$ & $1.54632$ & $1.54635$ & $0.00003$~~~~~~ \\
	  $ 7$ & $\nu_{1} + \nu_{2}$ & $123.25705$ & $123.25825$ & $9.8\times10^{-6}$ & $1.63610$ & $1.63641$ & $0.00031$~~~~~~ \\
	  $ 8$ & $\nu_{6}$ & $123.31830$ & $123.31967$ & $1.1\times10^{-5}$ & $1.69735$ & $1.69782$ & $0.00047$~~~~~~ \\
	  $ 9$ & $\nu_{7}$ & $123.45719$ & $123.45863$ & $1.2\times10^{-5}$ & $1.83624$ & $1.83679$ & $0.00055$~~~~~~ \\
	  $19$ & $2\nu_{1} + \nu_{2}$ & $123.92846$ & $123.92948$ & $8.3\times10^{-6}$ & $2.30751$ & $2.30764$ & $0.00013$~~~~~~ \\
	  $29$ & $\nu_{3} + \nu_{4}$ & $124.18986$ & $124.19103$ & $9.5\times10^{-6}$ & $2.56891$ & $2.56918$ & $0.00028$~~~~~~ \\
	  $39$ & $2\nu_{4}$ & $124.37948$ & $124.38091$ & $1.2\times10^{-5}$ & $2.75853$ & $2.75907$ & $0.00054$~~~~~~ \\
	  $49$ & $\nu_{2} + \nu_{8}$ & $124.55116$ & $124.55237$ & $9.7\times10^{-6}$ & $2.93021$ & $2.93052$ & $0.00032$~~~~~~ \\
	  $59$ & $\nu_{4} + \nu_{6}$ & $124.69756$ & $124.69933$ & $1.4\times10^{-5}$ & $3.07662$ & $3.07749$ & $0.00087$~~~~~~ \\
	  $69$ & $3\nu_{1} + \nu_{3}$ & $124.82482$ & $124.82576$ & $7.5\times10^{-6}$ & $3.20387$ & $3.20391$ & $0.00004$~~~~~~ \\
	  $79$ & $2\nu_{1} + \nu_{8}$ & $124.92929$ & $124.93030$ & $8.1\times10^{-6}$ & $3.30834$ & $3.30846$ & $0.00012$~~~~~ \\
		[0.5ex]
	  \hline\hline 
	\end{tabular}
	\label{Tab:64DCO}
\end{table*} 

Following this, we conducted a comprehensive measurement of $M=80$ energy levels. 
Owing to limitations in space, only a subset of these values is displayed in TABLE \ref{Tab:64DCO}.
A comparison with exact solutions reveals that the relative error ${(E_{n,\text{num}} - E_{n,\text{exact}})}/{E_{n,\text{exact}}}$ of our numerical results typically falls within the order of $10^{-6}$, while the absolute error $(E_{n,\text{num}} - E_{n,\text{exact}})$ spans from $0.001$ to $0.004$ energy units, demonstrating high precision in calculations.
In previous tensor-based methods, such RRBPM~\cite{leclerc2014calculating} and LOBPCG~\cite{rakhuba2016calculating},
have shown diminished accuracy for higher excited states, primarily due to the challenges posed by increased entanglement.
As mentioned in Refs.~\cite{thomas2015using,leclerc2014calculating},
the relative errors varying from $10^{-7}$ to $10^{-3}$.
We also compared the excitation energies by subtracting the ground state energy $\Delta E_{n,\text{num}} = E_{n,\text{num}} - E_{0,\text{num}}$, with errors consistently smaller than $0.001$ energy units.
In contrast, the variational wavefunctions are independent of entanglement in this method, 
ensuring consistent accuracy across all states,
while the equal weights of all states in the computation of the loss function.

It's noteworthy that utilizing the same RNVP network size $\{16, 128, 2\}$ to exclusively optimize the ground state wavefunction, 
i.e., setting the number of states $M=1$, yields more accurate results.
Under this condition, the ground state energy eventually converges to $E_{0,\text{num}}=121.620968(19)$ after $20000$ epochs. 
This represents a relative error of just $1.7\times 10^{-7}$ compared to the exact solution $E_{0,\text{exact}}=121.620948$, enhancing the precision by an order of magnitude compared to the ``ZPE" in TABLE \ref{Tab:64DCO}.
This demonstrates the superior accuracy of the neural canonical transformation method over the H-RRBPM~\cite{thomas2015using}.

In reality, the application of this method to a $64$-dimensional coupled oscillator may not fully demonstrate its unique advantages.
Because the square root of the Jacobian determinant, in Eq. (\ref{Eq:wavefunction}), plays a relatively simple role, which only serves to scale the coordinates (refer to Appendix \ref{sec:NDCO}) 
\marked{Despite this, as a benchmark test, it can be observed that our method does not rely on normal coordinates. This trivial transformation can be automatically achieved through the flow model, providing evidence of the reliability of the neural canonical transformation method in solving vibrational systems.}

\subsection{Acetonitrile}

\begin{figure}[h]
	\centering
	\includegraphics[width=0.45\textwidth]{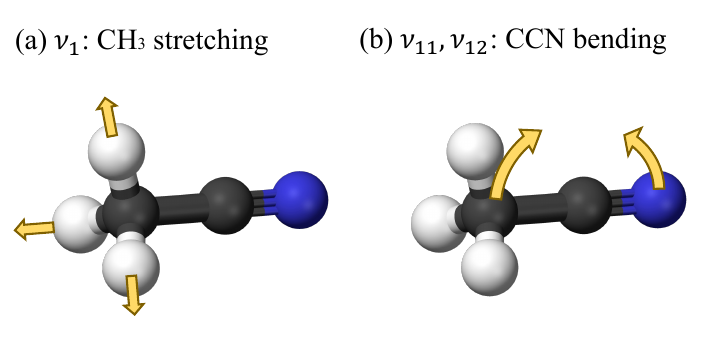}
	\caption{Two typical vibrational modes of acetonitrile ($\text{C}\text{H}_3\text{CN}$) with $C_{3v}$ symmetry. 
	The black, white and blue spheres represent $\text{C}$, $\text{H}$ and $\text{N}$ atoms respectively. 
	The vibrational mode $\nu_1$ has $\text{A}_1$ symmetry, 
	while mode $\nu_{11}$ and $\nu_{12}$ have $\text{E}$ symmetry.
	}
	\label{Fig:acetonitrile}
\end{figure}

Building on the discussion of the toy model in the preceding subsection,
we turn our attention to actual molecules, which exhibit multiple vibrational modes.
The exploration is critically important for this study, 
as it allows comparison with other numerical methods and experimental findings.
Specifically,
we have undertaken computational analyses on the acetonitrile ($\text{C}\text{H}_3\text{CN}$), 
a molecule that has been extensively studied~\cite{Herzberg1960MolecularSA,nakagawa1962rotation,Koivusaari1992high,tolonen1993the,andrews2000vibrational,paso1994the,duncan1978methyl,nakagawa1985infrared,begue2005calculations,thomas2015using,leclerc2014calculating,avila2011using,odunlami2017vci}.
It consists of 6 atoms and has $D=12$ vibrational modes, displaying $C_{3v}$ symmetry.
Using the same symbols as presented in Ref.~\cite{thomas2015using},
the vibrational modes $\nu_1,~\nu_2,~\nu_3,~\nu_4$ are characterized by $\text{A}_1$ symmetry.
Meanwhile, the modes $\nu_5,~\nu_7,~\nu_9,~\nu_{11}$ along with their degenerate counterparts $\nu_6,~\nu_8,~\nu_{10},~\nu_{12}$ are described by $\text{E}$ symmetry.
As illustrative instances,
FIG. \ref{Fig:acetonitrile} displays two characteristic vibrational modes, as elaborated in Ref.~\cite{begue2005calculations}.

In the following computing,
we used the same PES as described in the Refs.~\cite{thomas2015using,avila2011using,begue2005calculations},
which contains $12$ quadratic, $108$ cubic, and $191$ quartic terms, 
and it is publicly accessible in the code~\cite{mycode}. 
Despite the increased complexity of this PES, 
our method navigates the intricacies of such a vibrational Hamiltonian with proficiency. 
It is important to clarify that the approach does not necessitate a polynomial expansion of the PES. 
The decision to utilize the same force field was driven by our aim to enable direct, meaningful comparisons with the findings reported in Refs.~\cite{leclerc2014calculating,thomas2015using,avila2011using}.

\begin{table}[H]
	\caption{Comparison of Acetonitrile's Zero-Point Energies (ZPE) (in $\text{cm}^{-1}$) across various numerical methods.
    The neural canonical transformation method only focused on the ground state ($M=1$) with various sizes of RNVP networks.
	}
	\centering
	\begin{tabular}{L{4.5cm}L{3cm}}
	\hline\hline\\[-1.5ex]
	~~~~~~~~~~~~Method & ~~~~~~ZPE \\[0.5ex]
	\hline\\[-1.5ex]
	~~~~~~RRBPM\cite{leclerc2014calculating} & $ 9837.6293$ \\
	~~~~~~H-RRBPM~\cite{thomas2015using}  & $ 9837.429$ \\
	~~~~~~Smolyak~\cite{thomas2015using,avila2011using}  & $ 9837.4073$ \\[0.5ex]
	\hline\\[-1.5ex]
	~~~~~~RNVP \{16, 128, 2\} & $9837.40(2)$ \\
	~~~~~~RNVP \{16, 256, 2\} & $9837.42(3)$ \\
	~~~~~~RNVP \{32, 128, 2\} & $9837.39(3)$ \\
	~~~~~~RNVP \{64, 64, 2\}  & $9387.39(2)$ \\
	[0.5ex]
	\hline\hline
	\end{tabular}
	\label{Tab:CH3CNgroundstate}
\end{table}

At the beginning of this example,
our attention was dedicated to the optimization of the ground state wavefunction utilizing RNVP networks of various sizes.
We specified the number of states as $M=1$, explicitly indicating our objective to exclusively explore the ground state. 
The numerical results in the energy unit of wavenumbers ($\text{cm}^{-1}$) are presented in the TABLE \ref{Tab:CH3CNgroundstate},
and juxtaposed with findings from alternative numerical approaches~\cite{leclerc2014calculating,thomas2015using}.
For the purpose of our sampling was confined to the ground state,
we chose a sufficiently large training batch size of $65536$.

The comparison of energy values derived from all considered methods reveals a remarkable consistency, with discrepancies confined to the first decimal place and not exceeding $0.2~\text{cm}^{-1}$. 
We yields a ground state energy $E_0 = 9387.39(2)~\text{cm}^{-1}$ with the RNVP network $\{64, 64, 2\}$,
which is marginally lower than the energies obtained via the RRBPM and H-RRBPM, 
yet aligns closely with the outcomes produced by Smolyak's approach.
Notably, the precision of the energy estimations, reliant on Monte Carlo sampling, is inherently bound by the selected batch size.
As a result, enhancing accuracy and diminishing the margin of error present substantial challenges.

\begin{table*}[htp]
	\caption{Experimental and numerical energies (in $\text{cm}^{-1}$) of acetonitrile.
	The lowest $M=84$ states have been trained,
	and all computational results neglect contributions from \marked{vibrational angular momentum terms.}
	The excited state energies are adjusted relative to the ground state energy $E_n - E_0$.
	The Smolyak Quadrature Calculation results are labeled $E_\text{Smo}$ as the reference values.
	}
	\centering
	\begin{tabular}{L{1.4cm}L{3.4cm}R{2.0cm}R{2.0cm}R{2.0cm}R{2.0cm}R{2.0cm}R{2.0cm}}
		\hline\hline \\[-1.5ex]
		$n$ & Assignment & Experiment & Pert-Var & H-RRBPM & Smolyak & ~~~~~~~~~~~~~~~~~~~~~~~RNVP\{16,128,2\} \\
		& & \cite{nakagawa1962rotation,duncan1978methyl,nakagawa1985infrared,Koivusaari1992high,tolonen1993the,paso1994the,andrews2000vibrational}& 
		\cite{begue2005calculations} & \cite{thomas2015using} & \cite{thomas2015using,avila2011using}
		&$E$~~~~ &$E-E_\text{Smo}$ \\[0.8ex]
		\hline \\[-2ex]
		$ 1$&  $\nu_{11}$&  $362, 365$&  $366.0$&  $361.00$&  $360.991$&  $360.71$&  $-0.28$~~~~\\
		$ 3$&  $2\nu_{11}$&  $717 $&  $725.0$&  $723.22$&  $723.181$&  $723.44$&  $0.25$~~~~\\
		$ 4$&  $\nu_{11} + \nu_{12}$&  $739 $&  $731.0$&  $723.87$&  $723.827$&  $722.85$&  $-0.98$~~~~\\
		$ 6$&  $\nu_{4}$&  $916, 920 $&  $916.0$&  $900.71$&  $900.662$&  $898.83$&  $-1.83$~~~~\\
		$ 7$&  $\nu_{9}$&  $1041, 1042$&  $1038.0$&  $1034.19$&  $1034.126$&  $1034.56$&  $0.44$~~~~\\
		$ 9$&  $3\nu_{11}$&  $1122$&  $1098.0$&  $1086.66$&  $1086.554$&  $1087.94$&  $1.39$~~~~\\
		$12$&  $3\nu_{12}$&  $1122$&  $1098.0$&  $1086.66$&  $1086.554$&  $1087.81$&  $1.26$~~~~\\
		$10$&  $2\nu_{11} + \nu_{12}$&  $1077$&  $1094.0$&  $1087.88$&  $1087.776$&  $1086.66$&  $-1.11$~~~~\\
		$13$&  $\nu_{4} + \nu_{11}$&  $1290$&  $1282.0$&  $1259.90$&  $1259.822$&  $1257.97$&  $-1.85$~~~~\\
		$15$&  $\nu_{3}$&  $1390, 1385$&  $1400.0$&  $1389.16$&  $1388.973$&  $1396.64$&  $7.67$~~~~\\
		$16$&  $\nu_{9} + \nu_{11}$&  $1410, 1409$&  $1401.0$&  $1394.79$&  $1394.689$&  $1394.59$&  $-0.10$~~~~\\
		$17$&  $\nu_{10} + \nu_{11}$&  $1402$&  $1398.0$&  $1395.01$&  $1394.907$&  $1395.04$&  $0.13$~~~~\\
		$18$&  $\nu_{9} + \nu_{12}$&  $1402$&  $1398.0$&  $1397.85$&  $1397.687$&  $1395.25$&  $-2.44$~~~~\\
		$21$&  $3\nu_{11} + \nu_{12}$&  $1448$&  $1467.0$&  $1453.08$&  $1452.827$&  $1452.34$&  $-0.48$~~~~\\
		$25$&  $\nu_{7}$&  $1453, 1450$&  $1478.0$&  $1483.30$&  $1483.229$&  $1485.60$&  $2.37$~~~~\\
		$27$&  $\nu_{4} + 2\nu_{11}$&  $-$&  $1647.0$&  $1620.45$&  $1620.222$&  $1618.58$&  $-1.64$~~~~\\
		$28$&  $\nu_{4} + \nu_{11} + \nu_{12}$&  $-$&  $1645.7$&  $1620.99$&  $1620.767$&  $1618.03$&  $-2.73$~~~~\\
		$30$&  $\nu_{3} + \nu_{11}$&  $-$&  $1766.4$&  $1749.76$&  $1749.530$&  $1757.57$&  $8.04$~~~~\\
		$32$&  $\nu_{9} + 2\nu_{11}$&  $-$&  $1761.6$&  $1756.60$&  $1756.426$&  $1756.83$&  $0.40$~~~~\\
		$37$&  $\nu_{10} + 2\nu_{12}$&  $-$&  $1761.6$&  $1756.60$&  $1756.426$&  $1756.44$&  $0.02$~~~~\\
		$34$&  $\nu_{9} + \nu_{11} + \nu_{12}$&  $-$&  $1767.4$&  $1757.30$&  $1757.133$&  $1756.73$&  $-0.40$~~~~\\
		$33$&  $\nu_{10} + 2\nu_{11}$&  $-$&  $1769.3$&  $1759.98$&  $1759.772$&  $1757.50$&  $-2.27$~~~~\\
		$44$&  $2\nu_{4}$&  $-$&  $1833.7$&  $1785.37$&  $1785.207$&  $1792.26$&  $7.05$~~~~\\
		$46$&  $\nu_{8} + \nu_{11}$&  $-$&  $1838.2$&  $1844.36$&  $1844.258$&  $1845.96$&  $1.70$~~~~\\
		$47$&  $\nu_{7} + \nu_{12}$&  $-$&  $1838.2$&  $1844.79$&  $1844.690$&  $1846.48$&  $1.79$~~~~\\
		$45$&  $\nu_{7} + \nu_{11}$&  $-$&  $1842.2$&  $1844.43$&  $1844.330$&  $1846.58$&  $2.25$~~~~\\
		$49$&  $\nu_{4} + \nu_{9}$&  $-$&  $1952.3$&  $1931.70$&  $1931.547$&  $1932.51$&  $0.97$~~~~\\
		$51$&  $\nu_{4} + 3\nu_{11}$&  $-$&  $2010.3$&  $1982.14$&  $1981.849$&  $1981.05$&  $-0.80$~~~~\\
		$54$&  $\nu_{4} + 3\nu_{12}$&  $-$&  $2010.3$&  $1982.14$&  $1981.849$&  $1981.41$&  $-0.44$~~~~\\
		$52$&  $\nu_{4} + 2\nu_{11} + \nu_{12}$&  $-$&  $2015.1$&  $1983.14$&  $1982.857$&  $1980.05$&  $-2.81$~~~~\\
		$56$&  $\nu_{9} + \nu_{10}$&  $2075$&  $2059.0$&  $2057.17$&  $2057.068$&  $2065.40$&  $8.33$~~~~\\
		$55$&  $2\nu_{9}$&  $2082$&  $2067.0$&  $2065.38$&  $2065.286$&  $2063.61$&  $-1.68$~~~~\\
		$58$&  $\nu_{3} + 2\nu_{11}$&  $-$&  $2131.3$&  $2111.75$&  $2111.380$&  $2120.29$&  $8.91$~~~~\\
		$59$&  $\nu_{3} + \nu_{11} + \nu_{12}$&  $-$&  $2130.0$&  $2112.67$&  $2112.297$&  $2119.34$&  $7.05$~~~~\\
		$61$&  $\nu_{9} + 3\nu_{11}$&  $-$&  $2135.4$&  $2119.66$&  $2119.327$&  $2120.56$&  $1.23$~~~~\\
		$63$&  $\nu_{9} + 2\nu_{11} + \nu_{12}$&  $-$&  $2134.5$&  $2120.84$&  $2120.541$&  $2119.89$&  $-0.65$~~~~\\
		$64$&  $\nu_{10} + 2\nu_{11} + \nu_{12}$&  $-$&  $2130.5$&  $2121.21$&  $2120.910$&  $2120.02$&  $-0.89$~~~~\\
		$62$&  $\nu_{10} + 3\nu_{11}$&  $-$&  $2130.2$&  $2123.16$&  $2122.834$&  $2121.66$&  $-1.18$~~~~\\
		$65$&  $\nu_{9} + \nu_{11} + 2\nu_{12}$&  $-$&  $2130.5$&  $2123.63$&  $2123.301$&  $2120.14$&  $-3.16$~~~~\\
		$76$&  $2\nu_{4} + \nu_{11}$&  $-$&  $2199.4$&  $2142.68$&  $2142.614$&  $2149.64$&  $7.02$~~~~\\
		$78$&  $\nu_{7} + 2\nu_{11}$&  $2192$&  $2203.0$&  $2206.79$&  $2206.626$&  $2209.37$&  $2.74$~~~~\\
		$83$&  $\nu_{8} + 2\nu_{12}$&  $2192$&  $2203.0$&  $2206.80$&  $2206.633$&  $2208.96$&  $2.32$~~~~\\
		$79$&  $\nu_{8} + 2\nu_{11}$&  $-$&  $2210.4$&  $2206.95$&  $2206.776$&  $2208.56$&  $1.79$~~~~\\
		$80$&  $\nu_{7} + \nu_{11} + \nu_{12}$&  $-$&  $2208.3$&  $2207.72$&  $2207.559$&  $2208.49$&  $0.93$~~~~\\		
	    [0.5ex]
		\hline\hline 
	\end{tabular}\\
	\label{Tab:CH3CN}
\end{table*} 

Furthermore, this study extended to calculating the lowest $M=84$ excited states,
 using RNVP networks of various sizes.
To train a substantial number of states, 
we configured the batch size to $1024$,
resulting in the sampling of $1024\times 84$ states per epoch.
The corresponding numerical results along with the relevant experimental data~\cite{nakagawa1962rotation,duncan1978methyl,nakagawa1985infrared,Koivusaari1992high,tolonen1993the,paso1994the,andrews2000vibrational}
and numerical values~\cite{begue2005calculations,thomas2015using} 
are detailed in TABLE \ref{Tab:CH3CN}.
Training this network with batch size $B=1024$ and $M=84$ excited states for $10000$ epochs on a single A100 GPU takes approximately $13$ hours,
and the results typically converge well about $10000$ epochs for such systems.
It's noteworthy that the results for certain degenerate excited states are not listed due to their near-identical energy values, which fall within the confines of the $\text{E}$ symmetry group. 
By way of illustration, the energies for the excited states with assignment $\nu_{11}$ and $\nu_{12}$ are found to be $E_{\nu_{11}}=10199.52(8)~\text{cm}^{-1}$ and $E_{\nu_{12}}=10199.49(8)~\text{cm}^{-1}$, respectively, 
showcasing a negligible difference within the statistical error bar.

The energies of nearly all our excited states are lower than those reported in Ref.~\cite{begue2005calculations}, 
where the perturbation variational method was employed, 
underscoring the efficacy of the neural canonical transformation approach.
In comparison with results derived from tensor-based methods, 
such as Smolyak and H-RRBPM \cite{thomas2015using,avila2011using}, 
we have noticed some variations.
Although some energy levels are slightly higher and others lower, 
the differences generally fall within a narrow band of several wavenumbers.
Further analysis focusing on the layers of the RNVP network, 
including $\{16,128,2\}$ and $\{32,128,2\}$,
indicated no significant differences, 
leading us to exclude these results from the table.
We suspect that the disparities between our findings and the experimental results can be mainly attributed to the inherent limitations of the PES accuracy and the neglect of rotational terms. 
Because the other numerical techniques employed the same PES and disregarded the rotational terms as well.

\subsection{Ethylene oxide}\label{subsec:C2H4O}

\begin{figure}[htp]
	\centering
	\includegraphics[width=0.45\textwidth]{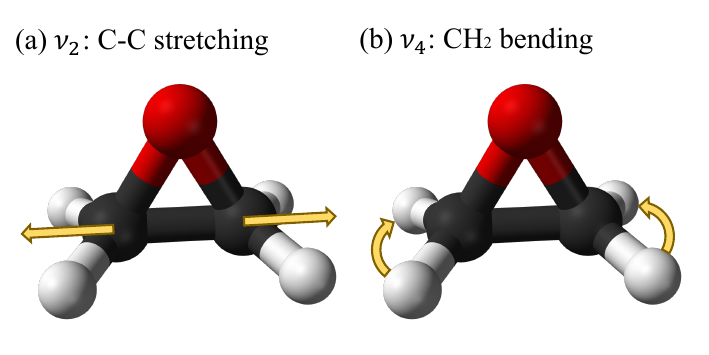}
	\caption{Two typical vibrational modes of ethylene oxide ($\text{C}_2\text{H}_4\text{O}$). 
	The black, white and red spheres represent $\text{C}$, $\text{H}$ and $\text{O}$ atoms respectively.
	}
	\label{Fig:EthyleneOxide}
\end{figure}

It is both motivating and rewarding for us to advance our computational investigations to a more complex molecule, specifically ethylene oxide ($\text{C}_2\text{H}_4\text{O}$),
which comprises $7$ atoms and possesses $15$ degrees of freedom.
As depicted in FIG. \ref{Fig:EthyleneOxide}, two typical vibrational modes are presented.
Unlike acetonitrile, ethylene oxide is distinguished by its lack of degenerate vibrational modes and exhibits a lower symmetry.
This molecule has been extensively studied through both experimental and computational techniques, 
as evidenced by a wealth of literature~\cite{lowe1986scaled,komornicki1983ab,rebsdat2000ethylene,begue2006single,cant1975vibrational,lord1956vibrational,schriver2004vibrational,nakanaga1980coriolis,yoshimizu1975microwave,carbonniere2010vci, zoccante2012approximate, thomas2015using}.
The same PES is taken in our calculation as described in Refs.~\cite{thomas2015using, begue2007comparison},
which comprises $15$ quadratic, $180$ cubic, and $445$ quartic terms~\cite{mycode}.

\begin{table}[H]
	\caption{Zero-point energies (ZPE)  (in $\text{cm}^{-1}$) of ethylene oxide from different numerical methods.
	The variational wavefunctions of the ground state ($M=1$) are optimized for various RNVP sizes. 
	}
	\centering
	\begin{tabular}{L{4cm}L{3cm}}
	\hline\hline\\[-1.5ex]
	~~~~~~~~~~~~Method & ~~~~~~ZPE \\[0.5ex]
	\hline\\[-1.5ex]
	~~~~~~VCI~\cite{begue2007comparison} & $12501$ \\
	~~~~~~VMFCI~\cite{begue2007comparison} & $12463.7$ \\
	~~~~~~VCI-P~\cite{carbonniere2010vci} & $12464.26$ \\
	~~~~~~H-RRBPM~\cite{thomas2015using}  & $12461.86$ \\[0.5ex]
	\hline\\[-1.5ex]
	~~~~~~RNVP \{16, 128, 2\} & $12461.59(4)$ \\
	~~~~~~RNVP \{16, 256, 2\} & $12461.63(5)$ \\
	~~~~~~RNVP \{32, 128, 2\} & $12461.58(4)$ \\
	~~~~~~RNVP \{64, 64, 2\}  & $12461.56(4)$ \\
	[0.5ex]
	\hline\hline
	\end{tabular}
	\label{Tab:C2H4Ogroundstate}
\end{table}

Similarly, we first train the neural network through the loss function Eq. (\ref{Eq:lossfunction}) with $M=1$,
employing a batch size of $65536$.
The results, detailed in TABLE \ref{Tab:C2H4Ogroundstate}, are compared with outcomes from other numerical methods as referenced in Refs.~\cite{begue2007comparison,carbonniere2010vci,thomas2015using}. 
The RNVP network with a size of $\{64, 64, 2\}$ achieves the lowest ground-state energy $E_0 = 12461.56~\text{cm}^{-1}$, 
surpassing the precision of other documented methods within the statistical error.
Subsequent investigations into the effect of RNVP network sizes, particularly those with $16$, $32$, and $64$ coupling layers, demonstrated a clear positive relationship between increased depth and lower ground-state energy.
However, enlarging the network's width showed a marginal decline in performance.
This phenomenon is linked to the number of network parameters, which exceeds 1.2 million in $\{16, 256, 2\}$, complicating convergence and requiring more epochs to converge.

\begin{table*}[htp]
	\caption{Experimental and numerical energies (in wavenumber $\text{cm}^{-1}$) of ethylene oxide.
	The number of states that have been trained in the loss function is $M=170$,
	and all numerical computational results in this table neglect contributions from \marked{vibrational angular momentum terms.}
	The excited state energies are corrected for the ground state background $ E_n - E_0 $.
	$^\text{a}$ The order of $n$ is taken from decoupled harmonic oscillators.
	$^\text{b}$ $\nu_6$ and $\nu_8$ are respectively sourced from gas phase in
	Ref.~\cite{cant1975vibrational} and Ref.~\cite{yoshimizu1975microwave}, 
	$\nu_7$ refers to crystalline phase~\cite{schriver2004vibrational},
	$2\nu_{10}$ and $2\nu_2$ are taken liquid phase~\cite{lord1956vibrational},
	$\nu_2+\nu_{10}$ is taken from solid phase~\cite{schriver2004vibrational},
	while all others are obtained from gas phase in Ref.~\cite{nakanaga1980coriolis}.
	$^\text{c}$ We use the same PES as VMFCI, VCI-P, and H-RRBPM, but differs from VCC.
	}
	\centering
	\begin{tabular}{L{2.0cm}L{2.8cm}R{2.0cm}R{2.0cm}R{2.0cm}R{2.0cm}R{2.0cm}R{2.0cm}}
		\hline\hline \\[-1.5ex]
		$n^\text{a}$ & Assignment & Experiment$^\text{b}$  
		& VMFCI & VCI-P & H-RRBPM & VCC$^\text{c}$ & RNVP~~~~  \\
		& & \cite{nakanaga1980coriolis,cant1975vibrational,yoshimizu1975microwave,schriver2004vibrational} & \cite{begue2007comparison} &
		\cite{carbonniere2010vci}& \cite{thomas2015using} &\cite{zoccante2012approximate}  &\{16,128,2\} \\[0.8ex]
		\hline \\[-2ex]
		$138$ & $\nu_{1}$ & $3018$  &$2919$ & $2921$ & $2920.93$ & $2956.6$ & $3006.50$   \\
		$11$ & $\nu_{2}$ & $1498$  &$1497$ & $1502$ & $1496.23$ & $1496.5$ & $1505.52$   \\
		$9$ & $\nu_{3}$ & $1270$  &$1272$ & $1275$ & $1271.12$ & $1267.0$ & $1270.05$   \\
		$6$ & $\nu_{4}$ & $1148$  &$1123$ & $1127$ & $1122.11$ & $1128.0$ & $1122.20$   \\
		$3$ & $\nu_{5}$ & $877$  &$879$ & $884$ & $878.40$ & $875.5$ & $875.44$   \\
		$163$ & $\nu_{6}$ & $3065$  &$3032$ & $3032$ & $3029.99$ & $3010.5$ & $3066.26$   \\
		$8$ & $\nu_{7}$ & $1045.8$  &$1149$ & $1154$ & $1148.64$ & $1151.7$ & $1145.09$   \\
		$4$ & $\nu_{8}$ & $1020$  &$1019$ & $1024$ & $1017.68$ & $1021.8$ & $1014.32$   \\
		$136$ & $\nu_{9}$ & $3006$  &$2913$ & $2929$ & $2912.48$ & $2899.5$ & $3001.21$   \\
		$10$ & $\nu_{10}$ & $1472$  &$1468$ & $1473$ & $1467.93$ & $1468.8$ & $1469.46$   \\
		$5$ & $\nu_{11}$ & $1151$  &$1125$ & $1131$ & $1124.65$ & $1122.9$ & $1122.58$   \\
		$2$ & $\nu_{12}$ & $840$  &$823$ & $827$ & $822.10$ & $818.9$ & $818.34$   \\
		$168$ & $\nu_{13}$ & $3065$  &$3041$ & $3043$ & $3041.79$ & $3021.3$ & $3080.13$   \\
		$7$ & $\nu_{14}$ & $1142$  &$1147$ & $1152$ & $1146.44$ & $1146.8$ & $1142.31$   \\
		$1$ & $\nu_{15}$ & $821$  &$794$ & $800$ & $793.29$ & $800.7$ & $788.01$   \\
		$70$ & $2\nu_{3}$ & $-$  &$-$ & $2545$ & $2537.64$ & $-$ & $2545.34$  \\
		$129$ & $\nu_{4} + \nu_{8} + \nu_{12}$ & $-$ &$-$ & $3032$ & $2940.11$ & $-$ & $2953.46$ \\
		$126$ & $\nu_{7} + \nu_{8} + \nu_{15}$ & $-$ &$-$ & $3053$ & $2951.19$ & $-$ & $2945.65$ \\
		$128$ & $\nu_{8} + \nu_{11} + \nu_{12}$ & $-$ &$-$ & $3056$ & $2953.39$ & $-$ & $2951.20$  \\
		$123$ & $2\nu_{10}$ & $2916$ &$-$ & $2963$ & $2960.45$ & $2918.0$ & $2938.59$ \\
		$130$ & $\nu_{2} + \nu_{10}$ & $2950.3$ &$-$ & $3032$ & $2998.85$ & $2990.3$ & $2976.43$   \\
		$133$ & $2\nu_{2}$ & $2961$ & $-$ & $3011$ & $3005.90$ & $3004.4$ & $3012.22$  \\
	    [0.5ex]
		\hline\hline 
	\end{tabular}\\
	\label{Tab:C2H4O}
\end{table*} 

In addition, 
we extended our computation to the lowest $M=170$ states across various RNVP network sizes, 
presenting a significantly more complex challenge than previous molecular studies. 
The experimental and computational data about these states are comprehensively detailed in TABLE \ref{Tab:C2H4O}.
The data are sourced from Refs.~\cite{cant1975vibrational,lord1956vibrational,schriver2004vibrational,nakanaga1980coriolis,yoshimizu1975microwave,carbonniere2010vci,zoccante2012approximate,thomas2015using},
where the VCC method employs distinct PES from other numerical calculations.
Training the network with architecture $\{16, 128, 2\}$ with batch size $B=1024$ and $M=170$ for $10000$ epochs on an A100 GPU requires approximately $25$ hours.

The results in TABLE \ref{Tab:C2H4O} indicate that for vibrational modes with energies below $1500 \text{cm}^{-1}$, the differences between our results and those obtained by other methods are minimal, generally within a few wave numbers.
For higher excited states, such as $2\nu_3$, $\nu_4+\nu_8+\nu_{12}$, as indicated in the bottom rows of the table, our results are in good agreement with the values obtained using the H-RRBPM method.
The primary discrepancies are observed in $\nu_1$ and $\nu_9$, where our method diverges more significantly from other numerical calculations.
In Ref.~\cite{begue2007comparison}, the authors observed convergence of $\nu_1$ and $\nu_9$ from about $3000\text{cm}^{-1}$ to the values in the table in both VMFCI and P-VMWCI methods.
Similarly, in the H-RRBPM method~\cite{thomas2015using}, significant variations of $\nu_1$ and $\nu_9$ under different parameters were also observed.
We notice that in the experiment, the four assignments $\nu_1$, $\nu_9$, $\nu_{6}$, $\nu_{13}$ are all above $3000\text{cm}^{-1}$ with no significant differences. 
However, the VMFCI method yields differences of around $100\text{cm}^{-1}$ between $\nu_1$, $\nu_9$, and $\nu_{6}$, $\nu_{13}$.
The authors~\cite{begue2007comparison} suspect this inconsistency arises from inaccuracies in the PES, which are already present at the harmonic level.

Since our approach utilizes the same PES as VMFCI, VCI-P, and H-RRBPM, we initially 
hypothesize that the source of difference in $\nu_1$ and $\nu_9$ may be attributed to the limited expressive capacity of the RNVP neural network.
Attempts to employ a deeper or wider RNVP network did not yield any significant changes, suggesting that a more complex network architecture might be necessary. 
Additionally, the choice of ADAM optimizer may be suboptimal, potentially causing convergence to local minima rather than the global optimum. 
We attempted to use a more complex stochastic reconfiguration optimizer~\cite{xie2023mstar,becca2017quantum}, 
but the results showed no difference.
Another possibility concerns the choice of the loss function, which is the sum of all energy levels.
During the training process, the loss function is minimized while ensuring the orthogonality of all states.
We utilized harmonic oscillator wavefunctions as the basis, and suspect that the choice of basis may not be optimal.
However, tests have shown that using VSCF wavefunctions as a basis does not improve the results.
It is also possible that our method exhibits larger errors for higher excited states. However, the results from the first model, $64$-dimensional coupled harmonic oscillator, show consistent accuracy across both low and high-excited states, without introducing significant errors in the higher states.

While both the VMFCI and H-RRBPM methods observe consistent convergence in $\nu_1$ and $\nu_9$, 
it is not entirely ruled out that their discrepancies with the experimental values are solely attributable to inaccuracies in the PES.
The VCC method, employing a different PES, aligns with VMFCI and H-RRBPM on $\nu_9$ but shows larger differences on $\nu_1$.
$\nu_1$, $\nu_9$, and $\nu_{6}$, $\nu_{13}$ also exhibit different trends in VCC compared to the experimental results.
Additionally, some other methods exhibit inconsistencies when computing higher excited states. 
For instance, the values of VCI-P for $\nu_1$ and $\nu_9$ are closer to those of H-RRBPM, 
but when computing the excitations $\nu_{4} + \nu_{8} + \nu_{12}$, $\nu_{7} + \nu_{8} + \nu_{15}$, and $\nu_{8} + \nu_{11} + \nu_{12}$, they are approximately $100\text{cm}^{-1}$ higher than H-RRBPM method.
In contrast, our results for these higher excited states are closer to those of H-RRBPM and VCC methods.
As mentioned earlier, the VMFCI method shows different trends for $\nu_1$, $\nu_9$ and $\nu_{6}$, $\nu_{13}$ compared to the experimental values.
Our method, using the same PES, exhibits significant differences in $\nu_1$, $\nu_9$, and $\nu_{6}$, $\nu_{13}$ at the harmonic level, but after computation, these values tend to converge, which is agree with the experimental values.

\section{Discussions}

Let's begin by comparing this approach with other related methods, especially the works related to machine learning.
\begin{itemize}

\item Similar to other methods such as VCI and VCC, our approach can be built upon the foundation of VSCF~\cite{bowman1978self,bowman1986self,gerber1979semiclassical,thompson1982optimization}. 
Initially, we can use the VSCF method to generate a new basis, 
and then use neural canonical transformation to optimize the wavefunctions. 
The wavefunctions derived from the VSCF method are more complex than those derived from the harmonic oscillators. 
However the testing showed that this approach leads to subtle improvements in the results, 
and it also significantly increases the computational time by two to three times with a $10$ level VSCF basis.

\item In Ref.~\cite{ishii2022development}, the authors use a backflow transformation into coordinate to solve the excited states of Hamiltonian, without an additional Jacobian determinant in the variational wavefunctions. 
They do not guarantee the orthogonality between different states, 
and the direct measurement of energies undertaken in their approach breaks the variational principle, making the results unreliable.

\item A recent study in Ref.~\cite{saleh2023computing} describes a normalizing flow parameterized by a residual neural network, and then combines the wavefunctions through superposition. 
However, due to the nonorthogonal wavefunctions,
their methodology requires a considerable amount of time for numerically integrating energy across multiple dimensions, which stands in contrast to the efficiency of our approach that uses Monte Carlo sampling for energy measurement.

\item When compared to another Ref.~\cite{pfau2023natural}, our method views atoms as a vibrational system and uses orthogonal wavefunctions, rather than constructing a nonorthogonal basis. 
This perspective enables us to avoid the complexities of electronic wavefunctions and diagonalizations associated with the number of states, significantly reducing computational complexity.
Whereas their method scales as $\mathcal{O}(M^4)$ for $M$ states, our method achieves linear scaling, $\mathcal{O}(M)$.

\item Our approach draws parallels with the ``Hermite+flow" methodology outlined in Ref.~\cite{cranmer2019inferring}, 
which employs a combination of Hermite polynomials and normalizing flow to represent the variational wavefunctions.
However, their analysis is restricted to single-particle transformations and does not extend to multi-particle systems.
In contrast, our research is adaptable to molecules with multiple atoms.

\end{itemize}
As a result, our approach can be extended to molecules with many degrees of freedom, making it a more scalable and versatile option.

In this research, the neural canonical transformations are start from normal coordinates, 
but in fact, it is not limited to normal coordinates.
The primary reason for the utilization of normal coordinates in this study, as opposed to Jacobi or Radau coordinates, is their ability to significantly simplify the kinetic energy operator.
Leveraging neural networks allows us to easily approximate the coordinate transformation between curvilinear and normal coordinates~\cite{Li2020}. 
Furthermore, it is worth noting that singularity is not a concern, as the transformation between normal and curvilinear coordinates remains reversible for practical computations.

Several aspects of our numerical calculations warrant further discussion.
The excited state wavefunctions of quantum harmonic oscillators have nodes,
which means we need a larger number of MCMC steps for effective sampling.
The reason behind this is that a large number of steps is essential to eliminate the correlation between consecutive samples, 
which is crucial for accurate simulations.
This requirement diverges from the systems existing in two or three dimensions without nodes,
as referred in Refs.~\cite{xie2023mstar,xie2023deep}.
During the training process, we implemented 200 MCMC steps to guarantee convergence.

The computational time required for this method is associated with multiple factors, which primarily comprise three parts: Monte Carlo sampling, the calculation of the loss function and its gradient, and the optimization with an optimizer. 
The ADAM optimizer we used does not essentially introduce additional cost, and over $90\%$ of the time is occupied by MCMC sampling and the computations related to the loss function. 
The computational time is principally linearly to the number of parameters in the neural network, the batch size, and the number of excited states.
Furthermore, employing $N$ GPUs for parallel computation requires only $1/N$ of the time, making it particularly suitable for parallel computing.

In Section \ref{sec:method}, for the sake of simplicity, we adopt an equal-weighted superposition approach within the loss function Eq. (\ref{Eq:lossfunction}).
Let us denote the weight associated with the $n$-th state during sampling as $w_n$, 
they must satisfy $w_0 \geq w_1 \geq w_2 \ldots \geq w_M$ and $w_i > 0$ 
by the variational principle~\cite{gross1988rayleigh}.
This arrangement offers flexibility in adjusting weights, 
which is crucial when dealing with a neural network with limited expressive capabilities. 
For instance, 
increasing the weight of the ground state can improve the accuracy of the ground state energy.
Furthermore, it is practical to study the system via a density matrix, 
$\rho = \sum_n \mu_n | \Psi_n(\bm{x})  \rangle  \langle \Psi_n(\bm{x}) |$, as mentioned in Refs.~ \cite{xie2022abinitio,xie2023mstar,cranmer2019inferring}.
Then introduce an extra hyperparameter, temperature, to adjust the probabilities of various states $\mu_n$.

This method presents significant opportunities for improvement, 
particularly in fully leveraging the capabilities of normalizing flows.
In theory, it can create a bijection between any probability distributions without changing the topological structure~\cite{papamakarios2021normalizing},
i.e. the nodes of the wavefunction in this research.
In the previous discussion, we mentioned that we can use a more powerful neural network to enhance the expressive ability.
For an enhanced neural network, several crucial attributes are necessary:
\begin{itemize}
\item The ability to perform beyond the linear transformations of RNVP's coupling layers, enabling non-linear transformations. 
\item A specialized Jacobian determinant structure, such as lower triangular or block-diagonal, facilitates faster computation.
\item The network needs to support second-order differentiation to avoid instability in kinetic energy computations.
\item Practical applications do not require the computation of inverse transformations, even though the neural canonical transformation theoretically needs to be bijective.
\end{itemize}
Given these considerations, several neural network models,
such as the neural spline flow model\cite{durkan2019neural} and autoregressive model~\cite{wu2019solving,liu2021solving,xie2023mstar}, 
have emerged as promising candidates for further exploration. 
It will be one of the key focuses of our future studies.

We have made all the source codes and original data publicly available to encourage further development~\cite{mycode}.
This method is notably scalable and requires no specific form of the PES.
This flexibility allows for integrating advanced computational approaches of the potential energies,
such as DFT and machine learning force field~\cite{lu2022fast,alvertis2023phonon}. 
These enhancements aim to improve potential energies beyond simple polynomials, 
a key area of our future research endeavors.
Moreover, leveraging machine learning strategies such as distributed training~\cite{goyal2017accurate},
enables this method to handle systems comprising hundreds of atoms efficiently. 
This capability is especially beneficial for studying complex molecular, 
including proteins.
We also plan to extend this approach to investigate the vibrational properties of solid systems~\cite{erba2019anharmonic}, focusing on phonons~\cite{monserrat2013anharmonic,monacelli2021stochastic},
as another main area of future research.
For such systems, we can use an autoregressive model~\cite{wu2019solving,liu2021solving,xie2023mstar} to encode a large number of states and then apply a flow model to parameterize the wavefunction.
Given these prospects, we are highly optimistic about the future evolution of neural canonical transformations in solving the vibrational spectra of vibrational systems.

\begin{acknowledgments}
We thank Tucker Carrington Jr. for providing the potential energy surface of ethylene oxide and acetonitrile. 
We are grateful for the useful discussions with Hao Xie, Zi-Hang Li, and Xing-Yu Zhang.
This work is supported by the National Natural Science Foundation of China under Grants No. T2225018, No. 92270107, No. 12188101, and No. T2121001, 
and the Strategic Priority Research Program of the Chinese Academy of Sciences under Grants No. XDB0500000 and No. XDB30000000.

\end{acknowledgments}

\section*{Data Availability Statement}

Data is openly available in a public repository that does not issue DOIs. The data that support the findings of this study are openly available in \url{https://github.com/zhangqi94/VibrationalSystem} and \url{https://github.com/zhangqi94/VibrationalSystemData}.

\section*{Conflict of Interest Statement}
The authors have no conflicts to disclose.

\appendix

\section{Basis state of variational wavefunction}\label{sec:basisstate}

In the variational wavefunctions Eq. (\ref{Eq:wavefunction}):
$
	\Psi_{n}(\bm{x})=\Phi_{n}(\bm{z})\left|\text{det}\left( {\partial \bm{z}}/{\partial \bm{x}} \right)\right|^{1/2},
$
we adopt the wavefunctions of a non-interacting harmonic oscillator in $D$ dimensions 
as the basis states.
It can be expressed as the direct product of a series of wavefunctions:
\begin{equation}\label{Eq:basisstate}
	\Phi_{n}(\bm{z})
	=\phi_{n_1}(z_1) ~\phi_{n_2}(z_2) ~... ~\phi_{n_D}(z_D),
\end{equation}
where $\phi_{n_i}(z_i)$ represents the eigenstates of a one-dimensional harmonic oscillator with frequency $\omega_i$:
\begin{equation}
	\phi_{n_i}(z_i) = 
	\exp \left(-\frac{1}{2}\omega_i z_i^2\right) 
	H_{n_i} \left(-\sqrt{{\omega_i}} z_i\right).
\end{equation}
In the above wavefunction, 
we have ignored the normalization factor,
$n_i$ is the energy level of the $i$-th mode,
and $H_{n_i}$ is the $n_i$-th order Hermite polynomial.
Then, the variational wavefunction is:
\begin{equation}
	\begin{aligned}
\Psi_{n}(x_1, x_2,..., x_D) &= \phi_{n_1}(z_1) ~\phi_{n_2}(z_2) ~... ~\phi_{n_D}(z_D)\\
	 &~~~~~~~~\times
	 \sqrt{\left|\text{det} \frac{\partial (z_1, z_2,..., z_D)}{\partial (x_1, x_2,..., x_D)} \right|}.
	\end{aligned}
\end{equation}

\section{The calculation of local energy}\label{sec:localenergy}

To ensure the numerical stability of local energy computing, 
it is essential to consider the logarithm of the wavefunction. 
The local energy with a given state $n$ and coordinate $\bm{x}$ is defined as:
\begin{equation}
	\begin{aligned}
  E_n^{\text{loc}}(\bm{x}) &= \frac{H \Psi_n(\bm{x})}{\Psi_n(\bm{x})} 
  = \frac{\left[ -\frac{1}{2} \frac{\partial^2}{\partial\bm{x}^2} + V(\bm{x})\right] \Psi_n(\bm{x})}{\Psi_n(\bm{x})} \\
  &= -\frac{1}{2} \frac{1}{\Psi_n(\bm{x})}\frac{\partial^2}{\partial\bm{x}^2}\Psi_n(\bm{x})+V(\bm{x}),
	\end{aligned}
\end{equation}
where the kinetic term is derived using the following transformations:
\begin{equation}
  \begin{aligned}
  \frac{1}{\Psi} \frac{\partial^2}{\partial\bm{x}^2} \Psi
  &= \frac{1}{\Psi} \frac{\partial}{\partial\bm{x}} \left( \frac{\partial}{\partial\bm{x}} \Psi \right) \\
  &= \frac{1}{\Psi} \frac{\partial}{\partial\bm{x}} \left( \Psi \frac{1}{\Psi} \frac{\partial}{\partial\bm{x}} \Psi \right) \\
  &= \frac{1}{\Psi} \frac{\partial}{\partial\bm{x}} \left( \Psi \frac{\partial}{\partial\bm{x}} \ln \Psi \right) \\
  &= \frac{1}{\Psi} \left[ \left( \frac{\partial}{\partial\bm{x}} \Psi \right) \left(\frac{\partial}{\partial\bm{x}} \ln \Psi\right) 
  + \Psi \frac{\partial^2}{\partial\bm{x}^2} \ln \Psi
  \right] \\
  &= \frac{1}{\Psi} \left( \frac{\partial}{\partial\bm{x}} \Psi \right) \left(\frac{\partial}{\partial\bm{x}} \ln \Psi\right) 
  + \frac{\partial^2}{\partial\bm{x}^2} \ln \Psi \\
  &= \left( \frac{\partial}{\partial\bm{x}} \ln \Psi \right)^2 + \frac{\partial^2}{\partial\bm{x}^2} \ln \Psi.
\end{aligned}
\end{equation}
Hence, the local energy can be expressed as Eq. (\ref{Eq:localenergy}):
\begin{equation}
	\begin{aligned}
	E_{n}^{\text{loc}}(\bm{x})
	=& - \frac{1}{2} \sum_{i=1}^D \left[
	\frac{\partial^2}{\partial x_i^2} \ln |\Psi_{n}(\bm{x})| 
	+ \left( \frac{\partial}{\partial x_i}  \ln |\Psi_{n}(\bm{x})|  \right)^2
	\right] \\&+ V(\bm{x}).
	\end{aligned}
\end{equation}

\section{Exact solution of D-dimensional coupled oscillator}\label{sec:NDCO}

As mentioned in the main text of the article Eq. (\ref{Eq:NDCO}),
the potential term of $D$-dimensional coupled harmonic oscillators is given by~\cite{rakhuba2016calculating,leclerc2014calculating,thomas2015using}:
\begin{equation}
	V = \sum_{i=1}^D \frac{1}{2} \omega_i^2 x_i^2 + \sum_{i>j}^{D} \alpha_{ij}\sqrt{\omega_i \omega_j} x_i x_j,
\end{equation}
where the coupling constant is taken as $\alpha_{ij}=\alpha~(i>j)$.
Also, the potential can be expressed in matrix form:
\begin{equation}
	V = \frac{1}{2} \bm{x}^T \bm{A} \bm{x},
\end{equation}
where $\bm{x} = (x_1, x_2, ..., x_D)^T$,
and $\bm{A}$ is a $D\times D$ positive-definite real-symmetric matrix with diagonal elements $\bm{A}_{ii}=\omega_i^2$ and off-diagonal elements $\bm{A}_{ij} = \alpha \sqrt{\omega_i \omega_j}$.
It could be exactly diagonalized as:
\begin{equation}
	\bm{A} = \bm{P} \bm{\mathnormal{\Lambda}} \bm{P}^{T},
\end{equation}
where $\bm{P}$ is a unitary matrix consisting of a series of eigenvectors,
which satisfied $\bm{P}^T \bm{P} = \bm{P} \bm{P}^T = \bm{I}$,
and $\bm{\mathnormal{\Lambda}} = \text{diag}(\nu_1^2, \nu_2^2, ... , \nu_D^2)$ is a diagonal matrix composed of eigenvalues. 
Taking the square root of these eigenvalues yields a set of new frequencies $\nu_i$. 
The exact solutions for the energy levels of $D$-dimensional coupled oscillators can be obtained by combining these frequencies:
\begin{equation}
	E_{n_1,n_2,...,n_D} = \sum_{i=1}^D \left(n_i+\frac{1}{2}\right)\nu_i.
\end{equation}

In Eq. (\ref{Eq:basisstate}), 
it is noted that the harmonic oscillator's frequency in the new coordinate $\bm{z}$ is $\omega_i$.
An additional step is required to incorporate a scaling operation to transform the frequency from $\nu_i$ to $\omega_i$.
Thus, the relationship between the new coordinates and the original coordinates involves a unitary transformation $\bm{P}^T$ and a scaling operation $\bm{\mathnormal{\Gamma}}$, i.e.:
\begin{equation}
	\bm{z} = \bm{\mathnormal{\Gamma}} \bm{P}^T \bm{x},
\end{equation}
where $\bm{\mathnormal{\Gamma}} = \text{diag}(\sqrt{\nu_1/\omega_1}, \sqrt{\nu_2/\omega_2}, ... , \sqrt{\nu_D/\omega_D}) $.
It is known that the Jacobian of unitary transformation $\bm{P}^T$ is a unit matrix,
hence the Jacobian determinant is only associated with the scaling operation:
\begin{equation}
	\left|\text{det}\left( \frac{\partial \bm{z}}{\partial \bm{x}} \right)\right| 
	= \prod_{i=1}^D \sqrt{\frac{\nu_i}{\omega_i}} .
\end{equation}

\section{Real-valued non-volume preserving neural network}\label{sec:RNVP}

It is essential to clarify that our goal is to parameterize the normalizing flow from the normal coordinates $\bm{x}$ to a set of new coordinates $\bm{z}$.
The well-known real-valued non-volume preserving (RNVP)~\cite{dinh2016density} is used to realize a normalizing flow used in generative models~\cite{wang2018generatemodels}. 
The transformation and jacobian determinant of each couping layer from $\bm{x}$ to $\bm{x'}$ is defined as follows:

\begin{enumerate}
\item Let $\bm{x} \in \mathcal{R}^D$ be a $D$ dimensional input coordinate, 
   and divide the input into two parts $\bm{x}_A,~\bm{x}_B$ with dimensions $d,~D-d$ respectively.

\item Denote the transform for a coupling layer as $f': \bm{x} \rightarrow \bm{x}'$, then the output follows the equations:
\begin{equation}
	\left\{
	\begin{aligned}
	& \bm{x}'_A = \bm{x}_A, \\
	& \bm{x}'_B = \bm{x}_B \odot \exp(s(\bm{x}_A)) + t(\bm{x}_A),
	\end{aligned}
	\right.
\end{equation}
where the symbol $\odot$ is the element-wise product.
The functions $s$ and $t$ correspond to scale and translation operations, 
and they are learnable mapping from $\mathcal{R}^{d}$ to $\mathcal{R}^{D-d}$.
These functions can be succinctly expressed using a multilayer perceptron (MLP).

\item The Jacobian determinant of the coupling layer transformation is given by a lower triangular matrix:
\begin{equation}
	\left| \frac{\partial \bm{x}'}{\partial \bm{x}} \right |
	= 
	\left|
	\begin{matrix}
	& \mathbb{I} & 0 \\
	& \frac{\partial{x}'_B}{\partial{x}_A} & \text{diag} [\exp(s(\bm{x}_A))] \\
	\end{matrix}
	\right|.
\end{equation}
Therefore, we can efficiently compute the logarithm of the determinant of this Jacobian without resorting to any derivative operations, i.e., 
$\log |\det ({\partial \bm{x}'}/{\partial \bm{x}})| = \sum_j s(\bm{x}_A)_j$.
\end{enumerate}

Repeated the above operations several times will create a comprehensive RNVP neural network,
capable of achieving a normalizing flow.
Each coupling layer in this context constitutes an invertible bijection, thereby making the entire network reversible.

\bibliography{mybibtex}

\end{document}